\documentclass[aps,prd,twocolumn,showpacs,amsmath,amssymb,nofootinbib]{revtex4}

\newcommand{\vague}[1]{`#1'}                                               
\newcommand{\spcpnct}{\;}                              
\newcommand{\period}{{\mbox{\spcpnct.}\relax}}         
\newcommand{\commae}{{\mbox{\spcpnct,}\relax}}         
\newcommand{\comma}{{\mbox{\spcpnct,\quad}\relax}}     

\newcommand{\tens}[1]{{\boldsymbol{#1}}}
\newcommand{\grad}{{\tens{d}}}

\newcommand{\covd}{{\tens{\nabla}}}
\newcommand{\pard}[1]{{\frac{\partial}{\partial #1}}}
\newcommand{\cv}[1]{{\tens{\partial}}_{#1}}
\newcommand{\ep}{{\tens{\epsilon}}}
\newcommand{\A}[1]{A^{\!(#1)}}

\newcommand{\detg}{{\mathfrak{g}}}

\newcommand{\qc}[1]{{k_{(#1)}}}
\newcommand{\lc}[1]{{l_{(#1)}}}
\newcommand{\qKT}[1]{{\tens{K}_{(#1)}}}
\newcommand{\lKT}[1]{{\tens{L}_{(#1)}}}
\newcommand{\qKTc}[1]{K_{(#1)}}

\newcommand{\qop}[1]{{\mathcal{K}_{(#1)}}}
\newcommand{\sop}[1]{{{\tilde{\mathcal{K}}}_{(#1)}}}
\newcommand{\lop}[1]{{\mathcal{L}_{(#1)}}}
\newcommand{\qsc}[1]{\Xi_{#1}}
\newcommand{\lsc}[1]{\Psi_{#1}}
\newcommand{\qsf}[1]{{\tilde\Xi}_{#1}}
\newcommand{\lsf}[1]{{\tilde\Psi}_{#1}}
\newcommand{\KGop}{\Box}

\newcommand{\scl}{\alpha}
\newcommand{\dd}{\varepsilon}

\newlength{\pri}
\begin{document}
\title{Complete Set of Commuting Symmetry Operators for the Klein-Gordon Equation in Generalized
Higher-Dimensional Kerr-NUT-(A)dS Spacetimes}

\author{Artur Sergyeyev}

\email{Artur.Sergyeyev@math.slu.cz}

\affiliation{Mathematical Institute, Silesian University in Opava,\\
Na Rybn\'\i{}\v{c}ku 1, 74601 Opava, Czech Republic}

\author{Pavel Krtou\v{s}}

\email{Pavel.Krtous@utf.mff.cuni.cz}

\affiliation{Institute of Theoretical Physics, Faculty of Mathematics and Physics, Charles University in Prague,\\
V~Hole\v{s}ovi\v{c}k\'ach 2, Prague, Czech Republic\medskip}

\date{January 7, 2008}  

\begin{abstract}
We consider the Klein-Gordon equation in generalized
higher-dimensional Kerr-NUT-(A)dS spacetime without imposing any
restrictions on the functional parameters characterizing the metric.
We establish commutativity of the second-order operators constructed
from the Killing tensors found in {[J. High Energy Phys. \textbf{02}
(2007) {004}]} and show that these operators, along with the
first-order operators originating from the Killing vectors, form a
complete set of commuting symmetry operators (i.e., integrals of
motion) for the Klein-Gordon equation. Moreover, we demonstrate that
the separated solutions of the Klein-Gordon equation obtained in
{[J. High Energy Phys. \textbf{02} (2007) {005}]} are joint
eigenfunctions for all of these operators. We also present explicit
form of the zero mode for the Klein-Gordon equation with zero mass.
\looseness=-1

In the semiclassical approximation we find that the
separated solutions of the Hamilton-Jacobi equation for geodesic
motion are also solutions for a set of Hamilton-Jacobi-type
equations which correspond to the quadratic conserved quantities
arising from the above Killing tensors.\looseness=-1
\end{abstract}


\maketitle


\section{Introduction}\label{sc:intro}
\vspace{-3mm} Investigation of the properties of higher-dimensional
black-hole spacetimes has recently attracted considerable attention,
in particular in connection with the string theory. The metrics
describing black holes of increasing generality were found in
\cite{Tangherlini:1963,MyersPerry:1986,HawkingEtal:1999,GibbonsEtal:2004,GibbonsEtal:2005}.
The most general metric of this kind known so far corresponds to a
higher-dimensional generally rotating (however neither charged nor
accelerated) black hole with the NUT parameters and arbitrary
cosmological constant. This metric was found by Chen, L\"u and Pope
\cite{ChenLuPope:2006} in the form which generalizes Carter's
four-dimensional Kerr-NUT-(anti-)de~Sitter metric
\cite{Carter:1968a,Carter:1968b}.\looseness=-1

The spacetime with the metric from \cite{ChenLuPope:2006} has a lot
of interesting properties. In ${D}$ dimensions it possesses explicit
and hidden symmetries encoded in the series of $n=[D/2]$ rank-two
Killing tensors and $D-n$ Killing vectors. The former ones can be
constructed from the so-called principal Killing-Yano tensor
\cite{KrtousEtal:2007a}, and in fact the spacetime in question is
the only one admitting a rank-two closed conformal Killing-Yano
tensor with certain further properties \cite{HouriEtal:2007}. The
symmetries allow one to define a complete set of ${D}$ quantities
conserved along geodesics. These quantities are linear and quadratic
in canonical momenta. Moreover, they are functionally independent
and in involution \cite{PageEtal:2007,KrtousEtal:2007b} and thus
their existence guarantees complete integrability of the geodesic
motion.\looseness=-1

The existence of such integrals of motion is intimately related to
separability of the Hamilton-Jacobi and Klein-Gor\-don equations. In
\cite{BenentiFrancaviglia:1979} it was shown that the presence of
these integrals yields the so-called separability structure. The
latter guarantees separability of the Hamilton-Jacobi equation and,
for the Einstein spaces, also separability of the Klein-Gordon
equation. Separability of the latter equation in the spacetime under
study was explicitly
demonstrated in \cite{FrolovEtal:2007}.\looseness=-1

In the present paper we discuss operator counterparts of the
conserved quantities constructed from the Killing vectors and
rank-two Killing tensors. Namely, we convert the integrals of motion
into operators using the rule ${\tens{p}\to -i\scl\covd}$ and
employing the symmetric ordering of derivatives, and we demonstrate
that all these operators commute. Since one of these operators is,
up to an an overall constant factor, 
the Klein-Gordon
operator itself, we thus obtain symmetry operators for the
Klein-Gordon equation in the sense of \cite{Miller:book1977,FushchichNikitin:book}.
Moreover, we show that the separated solutions of the Klein-Gordon
equation found in \cite{FrolovEtal:2007} are joint eigenfunctions of
all symmetry operators with eigenvalues corresponding to the
separation constants. As a byproduct, we obtain a zero mode solution
\eqref{zm} for the Klein-Gordon equation with zero
mass.\looseness=-1

We further demonstrate that semiclassical approximations of the
eigenvalue equations yield a set of Hamilton-Jacobi-type
equations. The latter can be solved using the separation of
variables in the same fashion as
in~\cite{FrolovEtal:2007}.\looseness=-1

It is worth noticing that all these properties actually hold for a
broader class of spacetimes than just the black-hole spacetimes of
\cite{ChenLuPope:2006}. These properties depend on the algebraic
structure of the metric \eqref{metric} rather than on the explicit
form of metric functions ${X_\mu}$. For this reason in what follows
we do not require our metric to satisfy the vacuum Einstein
equation.

\section{Preliminaries}

Consider $D$-dimensional spacetime with the metric
\begin{equation}\label{metric}
\begin{split}
\tens{g}
  =&\sum_{\mu=1}^n\;\biggl[\; \frac{U_\mu}{X_\mu}\,{\grad x_{\mu}^{\;\,2}}
  +\, \frac{X_\mu}{U_\mu}\,\Bigl(\,\sum_{j=0}^{n-1} \A{j}_{\mu}\grad\psi_j \Bigr)^{\!2}
  \;\biggr]\\
  &+\dd\frac{c}{\A{n}}\Bigl(\sum_{k=0}^n \A{k}\grad\psi_k\!\Bigr)^{\!2}\\
  =& \sum_{\mu=1}^n\,
    \biggl(\,\frac{U_\mu}{X_\mu}\,\ep^\mu  \ep^\mu
    + \frac{X_\mu}{U_\mu}\,\ep^{\hat \mu}  \ep^{\hat\mu}\,\biggr)
    + \dd\frac{c}{\A{n}}\,\ep^{\hat0} \ep^{\hat0}
  \period
\end{split}
\end{equation}
Here ${n=[D/2]}$, ${\dd=D-2n}$, and $c$ is an arbitrary constant;
${x_\mu}$, ${\mu=1,\dots,n}$, correspond to radial and latitudinal
directions while ${\psi_k}$, ${k=0,\dots,n+\dd-1}$, correspond to
temporal and longitudinal directions. The radial coordinate and some
other related quantities are actually rescaled by the imaginary unit
${i}$ in order to bring the metric into a more symmetric and compact
form, cf.\ e.g.~\cite{ChenLuPope:2006}. The signature of the metric
depends on the signs
of the metric functions; 
for the physically relevant ranges of coordinates it is
${(-+\dots+)}$.

We use Latin indices from the beginning of the alphabet to label the
whole coordinate set: ${\{x^a\}=\{x_\mu,\psi_k\}}$, where
${a=1,\dots,D}$, ${\mu=1,\dots,n}$, and ${k=0,\dots,n+\dd-1}$. The
non-normalized one-forms ${\{\ep_\mu,\ep_{\hat\mu}\}}$ that
diagonalize the metric, and the dual vector frame
${\{\ep^\mu,\ep^{\hat\mu}\}}$ have the form
\begin{gather}
  \mspace{-4mu}
  \ep^\mu = \grad x_{\mu}\comma
  \ep^{\hat\mu} = \sum_{j=0}^{n-1}\A{j}_{\mu}\grad\psi_j\comma
  \ep^{\hat0} = \sum_{k=0}^{n}\A{k}\grad\psi_k\commae\notag
\\[-1.5ex]
\label{frame}\\[-2ex]
  \mspace{-4mu}
  \ep_\mu = \cv{x_\mu}\!\comma\!\!\!
  \ep_{\hat\mu}\!=\!\!\!\!\sum_{k=0}^{n+\dd-1}\!\frac{(-x_{\mu}^2)^{n\!-\!1\!-\!k}}{U_{\mu}}\,\cv{\psi_{k}}\!\comma\!\!\!
  \ep_{\hat0}\!= \frac1{\A{n}}\cv{\psi_n}\!\period\!\!\!\notag
\end{gather}
The quantities $\ep^{\hat0}$ and $\ep_{\hat0}$ are defined only for
odd $D=2n+1$.
By ${\cv{x_\mu}}$ and ${\cv{\psi_k}}$ we denote the
coordinate vectors.

The functions ${U_\mu}$, ${\A{k}_\mu}$, ${U}$,
and ${\A{k}}$ used below are defined as follows
\begin{gather}
  \A{k}_{\mu}=\hspace{-5mm}\sum\limits_{\substack{\nu_1,\dots,\nu_k=1\\\nu_1<\dots<\nu_k,\;\nu_i\ne\mu}}^n\!\!\!\!\!
    x^2_{\nu_1}\cdots\ x^2_{\nu_k}\comma
  \A{k}=\!\!\!\!\!\sum\limits_{\substack{\nu_1,\dots,\nu_k=1\\\nu_1<\dots<\nu_k}}^n\!\!\!\!\!
    x^2_{\nu_1}\cdots\ x^2_{\nu_k}\commae\notag\\[-2ex]
  \label{UAdef}\\[-1ex]
  U_{\mu}=\prod\limits_{\substack{\nu=1\\\nu\ne\mu}}^{n}(x_{\nu}^2-x_{\mu}^2)\comma
  U=\prod\limits_{\substack{\mu,\nu=1\\\mu<\nu}}^n (x_\mu^2-x_\nu^2)=\det \A{j}_{\mu}\period\notag
\end{gather}
These functions satisfy the following important relations (see
e.g.~\cite{FrolovEtal:2007})
\begin{gather}
\hspace*{-1mm}  \sum_{\mu=1}^n\! \A{i}_\mu
\frac{(-x_\mu^2)^{n\!-\!1\!-\!j}}{U_\mu} =
  \delta^i_j\comma\!\!
  \sum_{j=0}^{n-1}\!\A{j}_\mu \frac{(-x_\nu^2)^{n\!-\!1\!-\!j}}{U_\nu} = \delta^\nu_\mu\commae  \notag\\[-1.5ex]
  \label{AUrel}\\[-1.5ex]
  \sum_{\mu=1}^n \frac{\A{j}_\mu}{x_\mu^2U_\mu} = \frac{\A{j}}{\A{n}}\comma
  \sum_{j=0}^n \A{j} {(-x_\mu^2)^{n\!-\!1\!-\!j}} = 0\commae\notag
\end{gather}
for ${i,j=0,\dots,n-1}$ and ${\mu,\nu=1,\dots,n}$. The determinant
of the metric in the coordinates ${\{x_\mu,\psi_k\}}$ reads
\begin{equation}\label{detgE}
  \detg=\det g_{ab} = (c\A{n})^\dd\, U^2\period
\end{equation}

In the above definitions we did not specify explicit form of the
functions~${X_\mu}$. In what follows we just assume that
${X_\mu=X_\mu(x_\mu)}$, ${\mu=1,\dots,n}$, i.e., each ${X_\mu}$ is a
function of a single variable ${x_\mu}$.
In general, the metric \eqref{metric} then does not satisfy the
vacuum Einstein equations. We would recover the vacuum black-hole
space\-time \cite{ChenLuPope:2006,HamamotoEtal:2007} by setting\looseness=-1
\begin{equation}\label{BHXs}
  X_\mu = b_\mu\, x_\mu^{1-\dd}-\dd\frac{ c}{x_\mu^2}
  + \sum_{k=\dd}^{n}\, c_{k}\, x_\mu^{2k}\period
\end{equation}
The constants $c$, ${c_k}$ and ${b_\mu}$ are then related to the
cosmological constant, angular momenta, mass, and NUT charges.

The spacetime with the metric \eqref{metric} possesses explicit
symmetries given by the Killing vectors
\begin{equation}\label{lKT}
  \lKT{k}=\cv{\psi_k}\comma {k=0,\dots,n+\dd-1}\commae
\end{equation}
and hidden symmetries that can be generated using the principal
Killing-Yano tensor \cite{KubiznakFrolov:2007,KrtousEtal:2007a}.
Namely, it was shown in \cite{KrtousEtal:2007a} that \eqref{metric}
admits a series of rank-two Killing tensors ${\qKT{j}}$,
${j=0,\dots,n-1}$ which are diagonal in the frame \eqref{frame} and
read
\begin{equation}\label{qKT}
\hspace*{-5pt}\qKT{j}\!=\!\sum_{\mu=1}^{n}\A{j}_\mu
  \biggl(\frac{X_\mu}{U_\mu}\,\ep_\mu   \ep_\mu
          + \frac{U_\mu}{X_\mu}\,\ep_{\hat \mu}\ep_{\hat \mu}\!\biggr)\!
  +\,\dd\A{j}\frac{\A{n}\!\!}{c}\,\ep_{\hat0} \ep_{\hat0}\!
  \period\!\!
\end{equation}

In particular, for ${j=0}$ we have ${\qKT{0}=\tens{g}^{-1}}$, where
$\tens{g}^{-1}$ is the contravariant metric. We use the Killing
tensors with contravariant indices as they are more convenient for
the construction of conserved quantities and symmetry operators used
below.

Let $\mathcal{M}$ be our spacetime with the metric \eqref{metric}
and $\mathcal{T}^*\mathcal{M}$ be the corresponding cotangent
bundle. The latter is naturally endowed with the canonical Poisson
bracket. Let ${\tens{p}}$ be the one-form of canonical momenta, so
that the components $p_a$ of ${\tens{p}}$ are canonically conjugate to
$x^a$, $a=1,\dots,D$. Then the above Killing vectors and tensors
generate conserved quantities for the geodesic motion on
$\mathcal{T}^*\mathcal{M}$.
The conserved quantities in question are
linear and quadratic in ${\tens{p}}$ and read
\begin{equation}\label{lqc}
\begin{aligned}
  \lc{k}&=\lKT{k}\cdot\tens{p}\comma k=0,\dots,n+\dd-1\commae\\
  \qc{j}&=\tens{p}\cdot\qKT{j}\cdot\tens{p}\comma j=0,\dots,n-1\period
\end{aligned}
\end{equation}
Here and below ${\cdot}$ denotes contraction.\looseness=-1
\pagebreak[2]

It was proved in
\cite{KrtousEtal:2007b} that these quantities are functionally
independent and in involution with respect to the canonical Poisson
bracket, and hence the geodesic motion on $\mathcal{T}^*\mathcal{M}$
is completely integrable.\looseness=-1

Recall that the geodesic motion in this context is generated by the
Hamiltonian
${H=\frac12\,\qc{0}=\frac12\,\tens{p}\cdot\tens{g}^{-1}\!\cdot\tens{p}}$,
and the fact that $\lc{k}$ and $\qc{j}$ are conserved quantities
(i.e., integrals) for the geodesic motion means that $\lc{k}$ and
$\qc{j}$ Poisson commute with $H$.

\section{Symmetry operators and separability for the Klein-Gordon equation}

It is natural to study the operators obtained from the above
conserved quantities using the heuristic rule ${\tens{p}\to
-i\scl\covd}$, where ${\covd}$ is the usual covariant derivative
with respect to the metric $\tens{g}$. Upon fixing the symmetric
operator ordering for the second order operators we
define
\begin{equation}\label{lqop}
\begin{aligned}
  \lop{k}&=-i\scl\,\lKT{k}\cdot\covd\commae\\
  \qop{j}&=-\scl^2\,\covd\cdot\bigl[\qKT{j}\cdot\covd\bigr]\period
\end{aligned}
\end{equation}
We employ here the convention that the square brackets do not
prevent the action of the derivatives to the right. Here ${\scl}$ is
a constant giving a scale to be used in order to obtain a
semiclassical (geometric-optical) approximation in the next section.
Of course, in the quantum context ${\scl}$ would be nothing but the
Planck constant ${\hbar}$. Writing out the above operators in the
coordinates ${\{x_\mu,\psi_k\}}$ we obtain\looseness=-1
\begin{equation}\label{lop}
  \lop{k}=-i\scl\,\pard{\psi_k}\comma k=0,\dots,n+\dd-1\commae
\end{equation}
\begin{equation}\label{qop}
\begin{split}
 \qop{j}=
    &-\scl^2\sum\limits_{\mu=1}^n\, \frac{\A{j}_\mu}{U_\mu} \Biggl[
      \pard{x_\mu} \biggl[
      X_\mu\pard{x_\mu}\biggr]+\dd\frac{X_\mu}{x_\mu}\pard{x_\mu}\\
    &\mspace{60mu}+\frac{1}{X_\mu}\biggl[\sum_{k=0}^{n+\dd-1}(-x_\mu^2)^{n\!-\!1\!-\!k}\pard{\psi_k}\biggr]^2
    \Biggr]\\
    &-\dd\frac{\scl^2\A{j}}{c\A{n}}\biggl[\pard{\psi_n}\biggr]^2
    \comma j=0,\dots,n-1
  \period
\end{split}
\end{equation}
Here we used the fact that $\qop{j}$ from \eqref{lqop} can be
written as
${\qop{j}=-\scl^2\,\detg^{\!-\!1/2}\partial_a\bigl[\detg^{\!1/2}\qKTc{j}^{ab}\,\partial_b\bigr]}$,
eq.~\eqref{detgE}, and a trivial identity
${\pard{x_\mu}\bigl(\A{j}_\mu U/U_\mu\bigr)=0}$.

In \cite{KrtousEtal:2007a} it was shown that the ``quasiclassical
limits" $\lc{k}$ and $\qc{j}$ of $\lop{k}$ and $\qop{j}$ are in
involution, i.e., they Poisson commute. However, the argument of
\cite{KrtousEtal:2007a} does not directly imply the commutativity of
the corresponding \vague{quantum} operators, i.e., of $\lop{k}$ and
$\qop{j}$.

Our goal is to show that the operators $\lop{k}$ and $\qop{j}$
commute for all ${k=0,\dots,n+\dd-1}$ and ${j=0,\dots,n-1}$.

The commutators of $\lop{k}$ among themselves and of $\lop{k}$ with
$\qop{j}$ obviously vanish, i.e.,
\begin{equation}\label{comm00}
  [\lop{k},\lop{l}]=0\comma
  [\lop{k},\qop{j}]=0
\end{equation}
for all ${k,l=0,\dots,n+\dd-1}$ and ${j=0,\dots,n-1}$, and we only
have to prove that
\begin{equation}\label{comm0}
   [\qop{i},\qop{j}]=0
\end{equation}
for all ${i,j=0,\dots,n-1}$.

To this end we first observe that
the operators $\qop{j}$ can be
written as
\begin{equation}\label{sop0}
  \qop{j} = \sum_{\mu=1}^n \frac{\A{j}_\mu}{U_\mu}\,\sop{\mu}\commae
\end{equation}
where
\begin{equation}\label{sopop}
\begin{split}
  \sop{\mu}=
      -\scl^2\Biggl[&\pard{x_\mu}\biggl[X_\mu\pard{x_\mu}\biggr]\!+\!\frac{\dd
      X_\mu}{x_\mu}\pard{x_\mu}+\frac{\dd}{c x_\mu^2}\,\biggl[\pard{\psi_n}\biggr]^2\\
      &+\frac{1}{X_\mu}\biggl[\sum_{k=0}^{n+\dd-1}\!\!(-x_\mu^2)^{n\!-\!1\!-\!k}\pard{\psi_k}\biggr]^2
      \Biggr]\period
\end{split}\raisetag{5ex}
\end{equation}
The operators \eqref{sopop} enjoy a remarkable property: for any given $\mu$
the operator $\sop{\mu}$ involves only $\partial/\partial x_\mu$ and
$x_\mu$ but does not involve $\partial/\partial x_\nu$ and $x_\nu$
for $\nu\neq\mu$. Therefore $\sop{\mu}$ commute, i.e., we have
\begin{equation}\label{comm1}
   [\sop{\mu},\sop{\nu}]=0
\end{equation}
for all $\mu,\nu=1,\dots,n$.

Using the identities \eqref{AUrel} we find that
\begin{equation}\label{sop}
  \sop{\mu}=\sum_{k=0}^{n-1}\,(-x_\mu^2)^{n\!-\!1\!-\!k}\,\qop{k}\commae
\end{equation}
and hence the commutativity of $\sop{\mu}$ entails that of $\qop{j}$.

Indeed, consider \eqref{comm1} for $\mu\neq \nu$ and rewrite it as
\begin{equation}\label{commwo}
\sop{\mu}\sop{\nu}=\sop{\nu}\sop{\mu}.
\end{equation}
Using \eqref{sop} for $\sop{\nu}$ on the left-hand side of
\eqref{commwo} and for $\sop{\mu}$ on the right-hand side of
\eqref{commwo} and employing a trivial identity
${[\sop{\mu},(-x_\nu^2)^{n-1-l}]=0}$ valid for ${\mu\neq\nu}$ yields\looseness=-1
\begin{equation}\label{comexp}
\hspace*{-5pt} \sum_{l=0}^{n-1}\,(-x_\nu^2)^{n\!-\!1\!-\!l} \sop{\mu} \qop{l} =
 \sum_{k=0}^{n-1}\,(-x_\mu^2)^{n\!-\!1\!-\!k} \sop{\nu} \qop{k}\period
\end{equation}
Again using \eqref{sop} we see that
\eqref{comexp} boils down to
\begin{equation}
  \sum_{k,l=0}^{n-1}\,
  (-x_\mu^2)^{n\!-\!1\!-\!k}
  (-x_\nu^2)^{n\!-\!1\!-l}\,
  [\qop{k},\qop{l}]=0\commae
\end{equation}
whence by non-singularity of the matrix
${B^j_\mu=(-x_\mu^2)^{n\!-\!1\!-\!j}}$ we readily obtain
\eqref{comm0}. It is important to stress that the above reasoning
makes substantial use of the fact that the matrix $B^j_\mu$ is the
St\"ackel matrix, i.e., its $\mu$-th column depends on $x_\mu$
alone, cf.\ e.g.\ \cite{KalninsMiller:1984}.

Notice an important corollary of \eqref{comm0}: since the Klein-Gordon equation
${\KGop\,\phi-\frac{m^2}{\scl^2}\phi=0}$
can be written as ${\qop{0}\phi=-m^2\phi}$, equations \eqref{comm00} and
\eqref{comm0} with ${i=0}$ imply that $\lop{k}$ and $\qop{j}$ are
\emph{symmetry operators} for the Klein-Gordon equation, see
e.g.~\cite{Miller:book1977,FushchichNikitin:book} and references
therein for the general theory of such operators.
Let us mention that for the special case of the Kerr
and Kerr--Newman metrics with $D=4$
this was established by Carter \cite{Carter:1977}.\looseness=-1

Relations \eqref{comm00} and \eqref{comm0} suggest that we may seek
for the joint eigenfunctions ${\phi}$ of the operators $\lop{k}$ and
$\qop{j}$ with the respective eigenvalues ${\lsc{k}}$ and
${\qsc{j}}$,
\begin{align}
  \lop{k}\phi &= \lsc{k}\,\phi\commae\label{lef}\\
  \qop{j}\phi &= \qsc{j}\,\phi\period\label{qef}
\end{align}

We will now show that these eigenfunctions can be found by
separation of variables, i.e., by assuming that they have the form
(see equation (4.2) of \cite{FrolovEtal:2007})
\begin{equation}\label{sepsolKG}
  \phi=\prod_{\mu=1}^{n}R_\mu(x_\mu) \;\prod_{k=0}^{n+\dd-1}
  \exp\Bigl(\,\frac{i}{\scl}\lsc{k}\psi_k\Bigr)\commae
\end{equation}
with each function ${R_\mu(x_\mu)}$ depending on a single variable
${x_\mu}$ only.

The functions \eqref{sepsolKG} clearly satisfy the equations
\eqref{lef}. Using \eqref{sop} we can
combine equations \eqref{qef} into an equivalent set of equations
\begin{equation}\label{sopeq}
  \biggl[\sop{\mu}-\sum_{j=0}^{n-1}\,\qsc{j}(-x_\mu^2)^{n\!-\!1\!-\!j}\biggr]\phi=0\period
\end{equation}
Note that these equations have the same form for different $\mu$,
i.e., the $\mu$-th equation \eqref{sopeq} is obtained from the
$\nu$-th one by replacing $x_\nu$ by $x_\mu$,  and vice versa.
Substituting \eqref{sopop} into \eqref{sopeq} we find that
\eqref{sepsolKG} solves all of the equations \eqref{sopeq}---and
therefore equations \eqref{qef} as well---provided the functions
${R_\mu}$ satisfy the following {\em ordinary} differential
equations:
\begin{equation}\label{sepeqnKG}
  \bigl(X_\mu R_\mu'\bigr)'
  +\dd \frac{X_\mu}{x_\mu}R_\mu'
  +\frac1{\scl^2}\Bigl(\qsf{\mu}-\frac{\lsf{\mu}^2}{X_\mu}\Bigr)R_\mu=0
  \commae
\end{equation}
where
\begin{equation}\label{OT}
\begin{split}
\hspace*{-10pt}
\lsf{\mu}\!\!=\!\!\!\sum_{k=0}^{n+\dd-1}\!\!\!\lsc{k}\,(-x_\mu^2)^{n\!-\!1\!-\!k}\comma\!\!
\qsf{\mu}\!\!=\!\!\!\sum_{k=0}^{n+\dd-1}\!\!\!\qsc{k}\,
(-x_\mu^2)^{n\!-\!1\!-\!k}\commae\!\!
\end{split}
\end{equation}
and for odd $D=2n+1$ we set ${\qsc{n}=\lsc{n}^2/c}$ for
convenience. Upon setting ${\scl}=1$ in these equations we recover,
of course, the separated ODEs (4.10) from \cite{FrolovEtal:2007}.

Thus, we proved that the functions ${\phi}$ of the form
\eqref{sepsolKG} with ${R_\mu}$ satisfying \eqref{sepeqnKG} are
eigenfunctions of the operators ${\lop{k}}$ and ${\qop{j}}$. In
particular, as ${\qop{0}=-\scl^2\KGop}$ (see above), we recover the
result of \cite{FrolovEtal:2007} that $\phi$ of the form
\eqref{sepsolKG} satisfies the Klein-Gordon equation
\begin{equation}\label{KGE}
\KGop\phi-{\displaystyle\frac{m^2}{\scl^2}}\,\phi=0\commae
\end{equation}
where $m^2=-\qsc{0}$.
\pagebreak[2]

As a final remark note that if $\lsc{k}=0$, $k=0,\dots,n+{\dd-1}$, and
$\qsc{j}=0$, $j=0,\dots,n-1$, then the general solution of
\eqref{sepeqnKG} is easily found to be
\begin{equation}\label{sepsolKGzms}
  R_\mu (x_\mu) = h_\mu+f_\mu\int \frac{dx_\mu}{X_\mu(x_\mu) x_\mu^{\dd}}\commae
\end{equation}
where $h_\mu$ and $f_\mu$ are arbitrary constants. Therefore,
\begin{equation}\label{zm}
\phi_0=\prod_{\mu=1}^{n}\left(h_\mu+f_\mu\int
\frac{dx_\mu}{X_\mu(x_\mu) x_\mu^{\dd}}\right)
\end{equation}
is a zero mode, i.e., it satisfies
\begin{equation}\label{zeromode}
  \lop{j}\phi_0=0\comma
  \qop{k}\phi_0=0
\end{equation}
for all $j=0,\dots, n+\dd-1$ and $k=0,\dots,n-1$.

\section{Hamilton-Jacobi equation}

Upon taking the solution ${\phi}$ of the Klein-Gordon equation in the form
\begin{equation}\label{semicl}
  \phi=A\exp\Bigl(\frac{i}{\scl}\,S\Bigr)
\end{equation}
we find that in the semiclassical (or geometric-optical, depending
on the application in question)
approximation 
the function ${S}$ satisfies the Hamilton-Jacobi equation
\begin{equation}\label{HJeq}
  \grad S \cdot \tens{g}^{-1} \cdot \grad S + m^2 =0\commae
\end{equation}
where ${m^2=-\qsc{0}}$. Recall that passing to the semiclassical
approximation in our case amounts to plugging the Ansatz \eqref{semicl} into
equation \eqref{KGE} multiplied by $\scl^2$
and taking the limit ${\scl\to 0}$.
\looseness=-1

The same approximation leads to the Hamilton-Jacobi-type equations
for each of the \vague{wave} equations \eqref{qef}, namely
\begin{equation}\label{gHJeq}
  \grad S \cdot \qKT{k} \cdot \grad S = \qsc{k}\period
\end{equation}

In a similar fashion, the quasiclassical limit of equations
\eqref{lef} yields
\begin{equation}\label{gLeq}
  \lKT{k}\cdot\grad S =\lsc{k}\period
\end{equation}
After the above discussion of the Klein-Gordon equation it is
not surprising that all these conditions are satisfied by the
separated solution found in \cite{FrolovEtal:2007}.

Indeed, the additive separation of
variables yields \cite{FrolovEtal:2007} the following Ansatz for ${S}$:
\begin{equation}\label{sepsolHJ}
  S=\sum_{\mu=1}^n S_\mu(x_\mu)+\sum_{k=0}^{n+\dd-1}\lsc{k}\psi_k\commae
\end{equation}
which automatically guarantees that the conditions \eqref{gLeq}
are satisfied.\looseness=-1

The Hamilton-Jacobi-type equations \eqref{gHJeq} lead
to the first-order differential equations
\begin{equation}\label{gHJsubst}
   \sum\limits_{\mu=1}^n\frac{\A{j}_\mu}{U_\mu}\biggl(
    X_\mu {S_\mu'}^2+\frac{\lsf{\mu}^2}{X_\mu}
    \biggr)
    +\dd \frac{\A{j}}{c\A{n}}\lsc{n}^2=\qsc{j}\period
\end{equation}
Taking linear combinations of these equations with the coefficients
given again by the entries of the matrix
${B^j_\mu=(-x_\mu^2)^{n\!-\!1\!-\!j}}$ yields an equivalent set of
ordinary differential equations for ${S_\mu}$'s,
\begin{equation}\label{sepeqHJ}
   {S_\mu'}^2= \frac{\qsf{\mu}}{X_\mu} -\frac{\lsf{\mu}^2}{X_\mu^2} \commae
\end{equation}
where $\lsf{\mu}$ and $\qsf{\mu}$ are given by \eqref{OT}
and we again set ${\qsc{n}=\lsc{n}^2/c}$ for odd $D=2n+1$.
Equations \eqref{sepeqHJ} are precisely the separated first-order
ODEs (3.7) of \cite{FrolovEtal:2007}.

By direct inspection we also find that equations \eqref{sepeqHJ}
yield the semiclassical approximation of the
separability conditions \eqref{sepeqnKG} for the Klein-Gordon
equation, with the functions ${R_\mu}$ related to ${S_\mu}$ as
${R_\mu=\exp(\frac{i}{\scl}S_\mu)}$.\looseness=-1

Finally, upon identifying the momentum vector ${\tens{p}=\grad S}$
in \eqref{gHJeq} and \eqref{gLeq} we recover the
original conserved quantities \eqref{lqc} in terms of the separation
constants:
\begin{equation}\label{lqclqsc}
  \lc{j}=\lsc{j}\comma \qc{j}=\qsc{j}\period
\end{equation}

\section{Conclusions and discussion}
In the present paper we have established that the quantities
${\lop{k}}$ and $\qop{j}$, see \eqref{lqop},
\eqref{lop} and \eqref{qop}, form a complete set of commuting
symmetry operators for the Klein-Gordon equation in the spacetime
with the metric \eqref{metric}. The symmetry operators ${\lop{k}}$
are associated with the Killing vectors, and the operators
${\qop{j}}$ with the Killing tensors constructed from the
Killing-Yano tensor \cite{KrtousEtal:2007a}. We proved the
commutativity of the symmetry operators in question for a general
class of metrics \eqref{metric} with each $X_\mu=X_\mu(x_\mu)$ being
an arbitrary function of a single variable $x_\mu$, $\mu=1,\dots,n$;
this class includes higher-dimensional Kerr-NUT-AdS metrics
\cite{ChenLuPope:2006} as special cases. We have further shown that
the separated solutions \eqref{sepsolKG} of the Klein-Gordon
equation found in \cite{FrolovEtal:2007} provide, cf.\
e.g.~\cite{KalninsMiller:1984,Miller:book1977}, joint eigenfunctions
of $\lop{k}$ and $\qop{j}$, see \eqref{lef} and \eqref{qef}. We have
also analyzed the quasiclassical limit of the above results and
compared this limit with the results of \cite{FrolovEtal:2007}.
Finally, we found explicit form of the zero mode \eqref{zm} for the
zero-mass Klein-Gordon equation.

In our opinion, it would be interesting to address similar issues
for the Dirac equation in the spacetime with the metric
\eqref{metric}, i.e., to find the symmetry operators whose joint
eigenfunctions are the separated solutions of the Dirac equation
found in \cite{OotaYasui:2007}.

\bigskip

\section*{Acknowledgments}
Both authors gratefully acknowledge the support from the Ministry of
Education, Youth and Sports of the Czech Republic under grants
MSM4781305904 (A.S.) and MSM0021610860 (P.K.). P.K. was also kindly
supported by the grant GA\v{C}R 202/08/0187. The authors are pleased
to thank the referee for useful suggestions. We also thank
D. Kubiz\v{n}\'ak for drawing our attention to the reference
\cite{Carter:1977}.



\begin{thebibliography}{10}


\bibitem{Tangherlini:1963}
F.~R. Tangherlini, Schwartzschild Field in $N$ Dimensions and the
  Dimensionality of Space Problem, Nuovo Cimento {\bf 27},  636  (1963).

\bibitem{MyersPerry:1986}
R.~C. Myers and M.~J. Perry, Black holes in higher dimensional
space-times, Ann. Phys. (N.Y.) {\bf 172},  304  (1986).

\bibitem{HawkingEtal:1999}
S.~W. Hawking, C.~J. Hunter, and M.~M. Taylor-Robinson, Rotation and
the  AdS/CFT correspondence, Phys. Rev. D {\bf 59},  064005 (1999),
 arXiv:hep-th/9811056.

\bibitem{GibbonsEtal:2004}
G.~W. Gibbons, H. L\"u, D.~N. Page, and C.~N. Pope, Rotating Black
Holes in  Higher Dimensions with a Cosmological Constant, Phys. Rev.
Lett. {\bf 93}, 171102  (2004), arXiv:hep-th/0409155.

\bibitem{GibbonsEtal:2005}
G.~W. Gibbons, H. L\"u, D.~N. Page, and C.~N. Pope, The General
Kerr-de Sitter  Metrics in All Dimensions, J. Geom. Phys. {\bf 53},
49  (2005),  arXiv:hep-th/0404008.

\bibitem{ChenLuPope:2006}
W. Chen, H. L\"u, and C.~N. Pope, General {Kerr-NUT-AdS} metrics in
all dimensions, Class. Quantum Grav. {\bf 23},  5323  (2006),
 arXiv:hep-th/0604125.

\bibitem{Carter:1968a}
B. Carter, A new family of Einstein spaces, Phys. Lett. {\bf 26A},
399 (1968).

\bibitem{Carter:1968b}
B. Carter, Hamilton-Jacobi and Schrodinger Separable Solutions of
Einstein's Equations, Commun. Math. Phys. {\bf 10},  280 (1968);
available online at
\url{http://projecteuclid.org/euclid.cmp/1103841118}.\looseness=-1

\bibitem{KrtousEtal:2007a}
P. Krtou\v{s}, D. Kubiz\v{n}\'ak, D.~N. Page, and V.~P. Frolov,
{Killing-Yano} tensors, rank-2 {Killing} tensors, and conserved
quantities in higher dimensions, J. High Energy Phys. {\bf 02}, 004
(2007), arXiv:hep-th/0612029.\looseness=-1

\bibitem{HouriEtal:2007}
T. Houri, T. Oota, and Y. Yasui, Closed conformal Killing-Yano
tensor and Kerr-NUT-de Sitter spacetime uniqueness, Phys. Lett. {\bf
B656},  214 (2007), arXiv:0708.1368 [hep-th].\looseness=-1

\bibitem{PageEtal:2007}
D.~N. Page, D. Kubiz\v{n}\'ak, M. Vasudevan, and P. Krtou\v{s},
Complete Integrability of Geodesic Motion in General
Higher-Di\-men\-sional Rotating Black Hole Spacetimes, Phys. Rev.
Lett. {\bf 98},  061102  (2007), arXiv:hep-th/0611083.\looseness=-1

\bibitem{KrtousEtal:2007b}
P. Krtou\v{s}, D. Kubiz\v{n}\'ak, D.~N. Page, and M. Vasudevan,
Constants of  Geodesic Motion in Higher-Dimensional Black-Hole
Spacetimes, Phys. Rev. D {\bf 76},  084034  (2007), arXiv:0707.0001
[hep-th].\looseness=-1

\bibitem{BenentiFrancaviglia:1979}
S. Benenti and M. Francaviglia, Remarks on certain separability
structures and their applications to general relativity, Gen. Rel.
Grav. {\bf 10},  79
  (1979).\looseness=-1

\bibitem{FrolovEtal:2007}
V.~P. Frolov, P. Krtou\v{s}, and D. Kubiz\v{n}\'ak, Separability of
  {Hamilton-Jacobi} and {Klein-Gordon} equations in general {Kerr-NUT-AdS}
  spacetimes, J. High Energy Phys. {\bf 02},  005  (2007),
  arXiv:hep-th/0611245.\looseness=-1

\bibitem{Miller:book1977}
W. Miller, Jr, {\em Symmetry and Separation of Variables}
(Addison-Wesley,
  Reading, Massachusetts, 1977);\\ available online at
\url{http://www.ima.umn.edu/~miller/}\linebreak\texttt{separationofvariables.html}.\looseness=-1

\bibitem{FushchichNikitin:book}
W.~I. Fushchich and A.~G. Nikitin, {\em Symmetry of Equations of
Quantum Mechanics} (Allerton, New York, 1994). \looseness=-1

\bibitem{HamamotoEtal:2007}
N. Hamamoto, T. Houri, T. Oota, and Y. Yasui, {Kerr-NUT-de~Sitter}
curvature in all dimensions, J. Phys. {\bf A40},  F177  (2007),
arXiv:hep-th/0611285. \looseness=-1

\bibitem{KubiznakFrolov:2007}
D. Kubiz\v{n}\'ak and V.~P. Frolov, Hidden symmetry of higher
dimensional {Kerr-NUT-AdS} spacetimes, Class. Quantum Grav. {\bf
24},  F1  (2007), arXiv:gr-qc/0610144. \looseness=-1


\bibitem{KalninsMiller:1984}
E.~G. Kalnins and W. Miller, Jr, The theory of orthogonal
R-separation for {Helmholtz} equations, Adv. Math. {\bf 51},  91
(1984); avaliable online at\\
\url{http://www.ima.umn.edu/~miller/Rsep.pdf}. \looseness=-1

\bibitem{Carter:1977}B.~Carter, Killing tensor quantum numbers
and conserved currents in curved space,
Phys. Rev. D \textbf{16}, 3395 (1977).

\bibitem{OotaYasui:2007}
T. Oota and Y. Yasui, Separability of the {Dirac} equation in higher
dimensional {Kerr--NUT--de Sitter} spacetime,   (2007),
arXiv:0711.0078  [hep-th]. \looseness=-1

\end{thebibliography}
\end{document}